\documentclass[preprint]{revtex4-2}

\usepackage{pstricks,graphicx}
\RequirePackage{mathbbol}
\RequirePackage{amssymb}
\RequirePackage{mathtools}

\usepackage{lmodern}
\usepackage{amsmath}
\usepackage{amsbsy}
\nonstopmode

\newcommand{\bra}[1]{\ensuremath{\langle #1|}}
\newcommand{\ket}[1]{\ensuremath{|#1\rangle}}

\newcommand{\braket}[2]{\ensuremath{\langle #1|#2\rangle}} 
\newcommand{\brakket}[1]{\ensuremath{\langle #1\rangle}} 
\newcommand{\dirint}[3]{\ensuremath{\langle #1|#2|#3\rangle}}

\newcommand{\bs}{\boldsymbol}
\newcommand{\lo}{\overline}
\newcommand{\half}{\frac{1}{2}}

\newcommand{\dul}[1]{\underline{\underline{#1}}}
\newcommand{\vtilde}[1]{\tilde{\raisebox{0pt}[0.85\height]{$\tilde{#1}$}}} 
\def\vcenterpdf#1{\mathrlap{\parbox{0pt}{\includegraphics[]{#1}}}\hphantom{\includegraphics[]{#1}}}
\begin{document}

\title{Non-interacting particle systems in the biorthogonal and unitary coupled-cluster description}
\author{J. Schirmer} 
\affiliation{Theoretische Chemie, Physikalisch-Chemisches Institut, Universit\"{a}t Heidelberg\\
Im Neuenheimer Feld 229, D-69120 Heidelberg, Germany}
\setcounter{tocdepth}{0}
\thispagestyle{empty}
\date{Oktober 26, 2023}

\begin{abstract}
Adapting a veritable many-body method to a system of non-interacting particles (NIP), while being trivial from a physical point of view, can be of interest with regard to methodological aspects. In this article we study the NIP versions of the biorthogonal (BCC) and unitary (UCC) coupled-cluster methods both for the ground state and generalized excitations. The essential simplification here is that the cluster operator is confined to single (or $p$-$h$) excitations.
For the ground state, specifically the CC amplitude equations, the NIP versions illustrate the enormous complexity of the UCC approach as compared to the BCC concept. In the treatment of excitations, on the other hand, the UCC outperforms the BCC method, as the UCC secular matrix is block-diagonal with respect to the excitation classes, e.g., $1h$ (one-hole), $2h$-$1p$ (two-hole-one particle), 
$\dots$ excitations of $N\!-\!1$ particles. Moreover, the UCC basis states are essentially equivalent with the states used in the ADC/ISR (algebraic-diagrammatic construction/intermediate state representation) approach, whose NIP version has been discussed in a recent paper [J. Schirmer, arXiv:2309.15721].
\end{abstract}

\maketitle

\section{Introduction}
In a recent paper~\cite{sch23:1} henceforth referred to as \textbf{I} an alternative approach to quasi-degenerate perturbation theory (QDPT) was presented that was obtained by simplifying established many-body techniques to systems of non-interacting particles (NIP). While the physics of a NIP system is trivial inasmuch as the solution of the underlying one-particle problem also provides  direct access to the many-particle ground and excited states, the NIP adaption of  many-body schemes may have interesting aspects at the formal level. 
As a striking example, elaborated in \textbf{I}, the ADC/ISR (algebraic-diagrammatic construction/intermediate-state representation) approach~\cite{sch98:4734,sch04:11449,Schirm:2018} to the particle-detachment part of the electron propagator  
turns into an unexpected new formalism of QDPT. But even apart from such a remarkable role change, there are some pedagogical benefits deriving from the NIP   transformations in that they can emphasize and clarify certain characteristics 
of the original many-body schemes.

The purpose of this paper is to present and analyze the NIP adaptions of the BCC (biorthogonal coupled-cluster) method~\cite{Helgaker:2000,Shavitt:2009}
and the UCC (unitary coupled-cluster) method~\cite{Muk:1989}, focussing here exemplarily on the treatment of ($N\!-\!1$)-particle excitations. Both CC variants have already been addressed in \textbf{I}, albeit in a very abbreviated form. The NIP form of the BCC method, for example, has been used for proving that the ISR secular matrix is block-diagonal with regard to the excitation classes, here, of one-hole ($1h$), two-hole-one-particle ($2h$-$1p$), $\dots$, excitations.  
As a supplement to \textbf{I}, a thorough and detailed treatment of ground and (generalized) excited states of a NIP system in the BCC and UCC framework will be presented in the following. In particular, a proof will be given of the finding stated in \textbf{I} that the ISR and UCC states are essentially equivalent or, at least in the lowest excitation class, even identical. 

The ensuing Sec.~II introduces the NIP concept and deals in detail with the associated BCC treatment. The UCC approach to NIP systems is presented in Sec.~III, together with a comparison with the ADC/ISR     formulation. Concluding remarks are given in Sec.~IV. There are three appendices. App.~A relates the Thouless expression for the NIP-BCC ground state amplitudes to the BCC amplitude equations. App.~B gives a proof of the block-diagonal structure of the NIP-UCC secular matrix. In App.~C it is shown that in the NIP case the unitary transformation relating the UCC and ISR states is block-diagonal as well.

\section{Biorthogonal coupled-cluster (BCC) approach to non-interacting particle (NIP) systems}
\subsection{Ground state}

In the following we consider a system of $N$ non-interacting fermionic particles, being subject to a hamiltonian of the form
\begin{equation}
\label{eq:Hmb}
\hat H = \hat H_0 + \hat W = \sum_p \epsilon_p c^\dagger_p c_p + \sum_{p,q} w_{pq}c^\dagger_p c_q
\end{equation}
where $\hat H_0$ is an 'unperturbed' hamiltonian and $\hat W$ is a one-particle operator. The fermion operators $c^\dagger_p, c_q$ refer to the basis of eigenfunctions $\phi_p (\xi), p= 1,2, \dots$ of the one-particle hamiltonian 
$\hat h_0$ underlying $\hat H_0$. Here the particle variables are denoted collectively by $\xi$, comprising, e.g., three spatial coordinates and a spin variable, $\xi \equiv \bs r\sigma$. 
The corresponding $N$-particle ground state is given by the Slater determinant 
\begin{equation}
\ket{\Phi_0} \equiv |\phi_1\phi_2\dots \phi_N|
\end{equation}

Using the basis set of $\phi_p$ functions the  one-particle Schr\"{o}dinger equation, 
\begin{equation} 
\label{eq:etp}
(\hat h_0 + \hat w) \psi_n(\bs r) = e_n \psi_n(\bs r), \;n = 1,2, \dots
\end{equation}
gives rise to the algebraic secular equation
\begin{equation}
\label{eq:ase}
(\bs \epsilon + \bs W) \bs X = \bs X \bs E
\end{equation}
Here, $\bs \epsilon$ is the diagonal matrix of the  
orbital energies $\epsilon_p$, and $\bs W$ is the matrix of the elements $w_{pq} = 
\dirint{\phi_p}{\hat w} {\phi_q}$. $\bs E$ and $\bs X$ denote the eigenvalue and eigenvector matrix, respectively.

The solutions of the one-particle problem~(\ref{eq:etp}) provide direct access to the exact ground state of the NIP system,
\begin{equation}
\label{eq:nipgs}
\ket{\Psi_0} \equiv |\psi_1\psi_2\dots \psi_N|
\end{equation}
as well as excited states, such as the singly excited states, 
\begin{equation}
\ket{\Psi_{ak}} = \tilde c^\dagger_a \tilde c_k \ket{\Psi_0}, \; 
 a > N, k \leq N 
\end{equation}
where the fermion operators $\tilde c^\dagger_n$ are associated with the eigenfunctions $\psi_n(\bs r)$ of $\hat h = \hat h_0 + \hat w$.
Here and in the following 
the quantum-chemical index notation is used, in which the indices 
$a,b,c,\dots$ refer to "unoccupied" orbitals ($> N$), $i,j,k, \dots$ refer to "occupied" orbitals ($\leq N$), while $p,q,r, \dots$ are unspecified.

In the BCC formulation the ground state of the NIP system is given by
\begin{equation}
\label{eq:bccgs}
\ket{\Psi'_0} = e^{\hat T_1}\ket{\Phi_0} = ( 1 + \hat T_1 + \half \hat T_1^2 + \frac{1}{6} \hat T_1^3 + \dots)
\ket{\Phi_0}
\end{equation}
where the $\hat T$ operator is restricted to single excitations,
\begin{equation}
\hat T = \hat T_1 = \sum t_{ak} c_a^\dagger c_k 
\end{equation}
Note that the BCC ground state supposes intermediate normalization, $\braket{\Phi_0}{\Psi'_0} = 1$, which is indicated by the prime superscript.

The BCC equations for the $p$-$h$ amplitudes $t_{ak}$ read
\begin{align}
\nonumber
0  =& \dirint{\Phi_{ak}}{e^{-\hat T_1} (\hat H_0 + \hat W) e^{\hat T_1}}{\Phi_0}\\
\label{eq: tak}
   =& \dirint{\Phi_{ak}}{(1 - \hat T_1)(\hat H_0 + \hat W) (1 + \hat T_1 + \half \hat T_1^2)}{\Phi_0}
\end{align} 
For obvious reasons, the first and second exponential expansion terminate after the linear and quadratic term, respectively. 
Altogether, there are two contributions associated with $\hat H_0$,
\begin{equation*}
\brakket{\hat H_0} \equiv -\dirint{\Phi_{ak}}{\hat T_1 \hat H_0}{\Phi_0} + \dirint{\Phi_{ak}}{\hat H_0 \hat T_1}{\Phi_0} = (\epsilon_a -\epsilon_k) t_{ak}
\end{equation*}
resulting in the expression on the right-hand side of the equation,
and 5 contributions with $\hat W$, 
\begin{align*}  
\brakket{\hat W} \equiv &\dirint{\Phi_{ak}}{\hat W}{\Phi_0} + \dirint{\Phi_{ak}}{\hat W \hat T_1}{\Phi_0}
- \dirint{\Phi_{ak}}{\hat T_1\hat W }{\Phi_0}\\
 +& \half \dirint{\Phi_{ak}}{\hat W \hat T_1^2}{\Phi_0} - \dirint{\Phi_{ak}}{\hat T_1 \hat W \hat T_1}{\Phi_0}
\end{align*}
Here, the terms 2 and 3 can be combined, as well as the terms 4 and 5, 
which, together with term 1, yields 
\begin{equation}
\brakket{\hat W} \equiv w_{ak} + \sum_b w_{ab}\, t_{bk}  - \sum_l t_{al}\,w_{lk} - \sum_{b,l}\, t_{al}\, w_{lb}\, t_{bk} 
\end{equation}
Note that in term 4 double excitations come into play, which makes the further evaluation somewhat tricky.
The resulting explicit BCC equations for the 
$t_{ak}$ amplitudes read 
\begin{equation}
\label{eq:bccamp}                                                                                       
t_{ak} = - \frac{1}{\epsilon_a -\epsilon_k} ( w_{ak} + \sum_b w_{ab}\,t_{bk}  - \sum_l t_{al}\,w_{lk} - \sum_{b,l} t_{al} \,w_{lb}\, t_{bk})
\end{equation}
These are implicit quadratic equations for the amplitudes to be solved by means of iteration. They can also be evaluated using perturbation theory in a successive way. The first three orders are given by 
\begin{align}
\nonumber
t^{(1)}_{ak} &= -\frac{w_{ak}}{\epsilon_a -\epsilon_k}\\ 
\nonumber
t^{(2)}_{ak} &= -\frac{1}{\epsilon_a -\epsilon_k} ( \sum_b w_{ab}\,t^{(1)}_{bk} - \sum_l t^{(1)}_{al}w_{lk})\\
\label{eq:bccpt}
t^{(3)}_{ak} &= -\frac{1}{\epsilon_a -\epsilon_k} ( \sum_b w_{ab}\,t^{(2)}_{bk} - \sum_l t^{(2)}_{al}w_{lk} )
+ \frac{1}{\epsilon_a -\epsilon_k} \sum_{b,l} t^{(1)}_{al} w_{lb}\, t^{(1)}_{bk} 
\end{align}
The PT expansions can also be obtained by resorting to the usual perturbation treatment of the NIP ground state in the standard CI representation,
\begin{equation}
\label{eq:cigs}
\ket{\Psi'_0} = (1 +\hat X_1 + \hat X_2 + \hat X_3 + \dots)\ket{\Phi_0}
\end{equation}
where $\hat X_\nu$ are the class-specific excitation operators, such as the operator for the $p$-$h$ excitations, 
\begin{equation}
\hat X_1 = \sum X_{ak}c_a^\dagger c_k
\end{equation}
As the comparison of the CI expansion with the BCC form~(\ref{eq:bccgs}) shows,
\begin{equation*}
\hat X_1 = \hat T_1, \;\; 
\hat X_2 = 1/2 \,\hat T_1^2, \;\;
\hat X_3 = 1/6\,\hat T_1^3, \; \dots 
\end{equation*}

A closed-form expression for the amplitudes in the NIP-BCC ground state in terms of eigenvector components of the underlying one-particle problem~(\ref{eq:etp}) is provided by a theorem of Thouless~\cite{tho60:225}.   
Let $\bs X$ denote the matrix of eigenvectors of Eq.~(\ref{eq:ase}) and consider
a partitioning of $\bs X$ according to
\begin{equation}
\label{eq:part0}
\bs X =
\left(
\begin{array}{cc}
\bs X_{11} &\bs X_{12} \\
\bs X_{21} &\bs X_{22}
\end{array}
\right)
\end{equation}
Here, the subscripts $1, 2$ in the second position refer to eigenvectors $\bs x_n, n \leq N$ and $\bs x_n, n > N$, respectively, while in the first position they distinguish the basis functions $\phi_p$ according to $p \leq N$ 
and $p > N$, respectively.   
Now the Thouless expression can be written as 
\begin{equation}
\label{eq:t21}
\bs t_{21} = \bs X_{21} \bs X_{11}^{-1}
\end{equation}
where $\bs t_{21}$ denotes the matrix of the BCC amplitudes $t_{ak}$.
This result is obtained by applying Thouless' theorem to the exact NIP ground state~(\ref{eq:nipgs}). It should be of interest to verify that the Thouless expression~(\ref{eq:t21}) is indeed a solution of the BCC amplitude equations~(\ref{eq:bccgs}). This is demonstrated in App. A.

\subsection{Excited states}

The BCC approach to general excited states of a NIP system has already been discussed at some detail in \textbf{I}, so that here we may confine us to a brief sketch.
As in \textbf{I} we will specifically deal with excitations in the ($N\!-\!1$)-particle system (particle removal), which may be seen as being exemplary for other cases, such as $N$-electron excitations, particle attachment, etc. 

The BCC treatment of ($N\!-\!1$)-particle excitations is based on the 
non-hermitian representation of the 
(shifted) hamiltonian $\hat{H}-E^N_{0}$, 
\begin{equation}
\label{eq:ccsm}
M^{bcc}_{IJ}=\bra{\lo{\Phi}_{I}}\hat{H}-E^N_{0}\ket{\Psi_{J}^{0}}
= \dirint{\Phi}{\hat{C}_{I}^{\dagger} e^{-\hat{T}_1}
[\hat{H},\hat{C}_{J}] e^{\hat{T}_1}}{\Phi}.
\end{equation}
in terms of two distinct sets of ($N\!-\!1$)-particle states.
There is a right expansion manifold $\{R\}$ formed by
the correlated excited states,
\begin{equation}
\label{eq:ccstates}
\ket{\Psi^0_{J}}=
\hat{C}_{J} \ket{\Psi^{cc}_0}=
\hat{C}_{J} e^{\hat{T}_1}\ket{\Phi} =  e^{\hat{T}_1}\hat{C}_{J} \ket{\Phi}
\end{equation}
where $\hat C_J$ denote physical excitation operators of the manifold of $1h$, $2h$-$1p$, $\dots$, excitations,
\begin{equation}
\label{eq:bccmf}
\{\hat C_J\} = \{c_k; c_a^\dagger c_k c_l, k<l; \dots\}
\end{equation}
The left expansion manifold $\{L\}$ is formed by the associated biorthogonal states, 
\begin{equation}
\label{eq:bostates}
\bra{\lo{\Phi}_{I}}=
\bra{\Phi}\hat{C}_{I}^{\dagger} e^{-\hat{T}_1}
\end{equation}
According to the two expansion sets, there is a left and a right eigenvalue problem, giving rise to the same excitation (particle removal) energies.

\begin{figure}
\begin{equation*}
\vcenterpdf{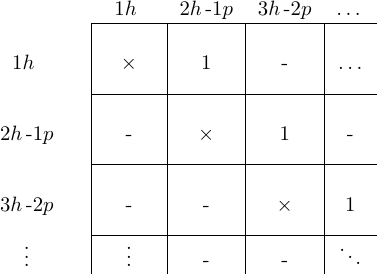}
\end{equation*}
\caption{Block structure of the biorthogonal coupled-cluster (BCC) secular matrix $\bs M^{bcc}$ for non-interacting particles. The entry "1" means (first-order) coupling via the perturbation part $\hat W$ of the many-particle hamiltonian.}
\label{fig:bccorfigx}
\end{figure}

As analyzed in \textbf{I}, the BCC secular matrix $\bs M^{bcc}$ has the block structure shown in Fig.~\ref{fig:bccorfigx}. In the upper right part there is a  
CI-type structure featuring a coupling between each two successive excitation classes, whereas the lower left part is block-diagonal, that is, the excitation classes are entirely decoupled from each other.  
In the UCC approach to be discussed in the ensuing section the correspnding  secular matrix is hermitian and block-diagonal.

\section{NIP systems in the unitary coupled-cluster (UCC) formulation}
\subsection{Ground state}

The UCC representation of the NIP ground state~(\ref{eq:nipgs}) is given by 
\begin{equation}
\label{eq:uccgs}
\ket{\Psi_0} = e^{\hat \sigma}\ket{\Phi_0} = ( 1 + \hat S_1 + \half \{\hat S_1^2 - \hat S_1^\dagger \hat S_1\}  + 
\frac{1}{6} \{\hat S_1^3 -\hat S_1 \hat S_1^\dagger \hat S_1 - \hat S_1^\dagger \hat S_1^2 \} + \dots)\ket{\Phi_0}
\end{equation}
where the anti-hermitian UCC excitation operator $\hat \sigma$ is restricted to single excitations,
\begin{equation}
\label{eq:siop}
\hat \sigma = \hat S_1 - \hat S_1^\dagger, \; \;\hat S_1 = \sum S_{ak} c_a^\dagger c_k 
\end{equation}
In the UCC form the ground state is normalized to 1,
that is, the UCC ground state~(\ref{eq:uccgs}) and the BCC ground state~(\ref{eq:bccgs}) 
differ by a normalization factor,
\begin{equation}
\label{eq:ubcc}
\ket{\Psi_0} = I_0^{-1/2} \ket{\Psi'_0}
\end{equation}
where $I_0$ is the BCC normalization integral
\begin{equation}
\label{eq:norm}
I_0 = \braket{\Psi'_0}{\Psi'_0} = \dirint{\Phi_0}{(e^{\hat T_1})^\dagger e^{\hat T_1}}{\Phi_0}
\end{equation} 

The UCC amplitude equations read
\begin{equation}
\label{eq:uccamp}
\dirint{\Phi_{ak}}{e^{-\hat \sigma}(\hat H_0 + \hat W)e^{\hat \sigma}}{\Phi_0} = 0
\end{equation}
Unlike the case of BCC, they do not terminate after a finite number of terms in the exponential expansions, which indicates a substantial complication for their computational evaluation.
The complexity of the UCC amplitude equations becomes apparent already at low-order perturbation theory for the amplitudes,
and we shall take a closer look at that in the following. 

Let us first consider the $p$-$h$ amplitudes 
\begin{equation}
\label{eq:xxak}
x_{ak} = \braket{\Phi_{ak}}{\Psi_0} 
\end{equation}
in the NIP ground state in the UCC representation~(\ref{eq:uccgs}), which through second order are identical with
the CI amplitudes $X_{ak}$ (Eq.~\ref{eq:cigs}) and the BCC amplitudes
$t_{ak}$ pertaining to the case of intermediate normalization,
\[X^{(n)}_{ak} = t^{(n)}_{ak} =  x^{(n)}_{ak}, \; n = 1,2\]
While in first and second order such relations apply also to the UCC amplitudes $S_{ak}$,
\begin{equation}
x_{ak}^{(1)} = S_{ak}^{(1)}, \;\; x_{ak}^{(2)} = S_{ak}^{(2)} 
\end{equation} 
things become way more intricate in third order, where already cubic terms of the exponential expansion come into play:
\begin{equation}
\label{eq:xak3}
x_{ak}^{(3)} = S_{ak}^{(3)} - \frac{1}{6} \dirint{\Phi_{ak}}{\hat S_1^{(1)} \hat S_1^{(1)\dagger} \hat S_1^{(1)} }{\Phi_0} 
- \frac{1}{6} \dirint{\Phi_{ak}}{\hat S_1^{(1)\dagger} \hat S_1^{(1)\,2}}{\Phi_0} 
\end{equation} 
As indicated by this expression, the full amplitude $x_{ak}$ will be constituted in a complicated way by contributions from any order of the exponential expansion.

The perturbation expansion of the UCC amplitudes $S_{ak}$, beginning in first order, can obtained successively from the amplitude equations~(\ref{eq:uccamp}). In first and second order, this reproduces just the corresponding BCC terms~(\ref{eq:bccpt}), 
$S^{(1)}_{ak} = t^{(1)}_{ak}$ and $S^{(2)}_{ak} = t^{(2)}_{ak}$.

The third-order contribution is given by
\begin{equation}
\label{eq:uccpt}
S^{(3)}_{ak} = t^{(3)}_{ak} + \frac{1}{6} \dirint{\Phi_{ak}}{\hat S_1^{(1)} \hat S_1^{(1)\dagger} \hat S_1^{(1)} }{\Phi_0} 
+ \frac{1}{6} \dirint{\Phi_{ak}}{\hat S_1^{(1)\dagger} \hat S_1^{(1)\,2}}{\Phi_0} 
- \half I_0^{(2)} S_{ak}^{(1)}
\end{equation}
where $t^{(3)}_{ak}$ is the third-order BCC amplitude according to 
Eq.~(\ref{eq:bccpt}) and $I_0^{(2)}$ 
is the second-order contribution to the normalization integral~(\ref{eq:norm}). 
The evaluation of the latter result is already somewhat tedious.
Both the expansions of $e^{\hat \sigma}\ket{\Phi_0}$ and $\bra{\Phi_{ak}}e^{- \hat \sigma}$ in the amplitude equations are needed through the cubic terms. Altogether there are 18 non-vanishing third-order contributions, of which 10 relate to 
$\hat H_0$ and 8 to $\hat W$. The first 6 $\hat H_0\,$ terms give rise to 
$S^{(3)}_{ak}$ on the left side of Eq.~(\ref{eq:uccpt}) plus the second and third term on the right-hand side. Four of the $\hat W$ terms reproduce the 4 contributions to $t_{ak}^{(3)}$, another one  yields the normalization contribution, $-\half \, I_0^{(2)} S_{ak}^{(1)}$. There remain 4 terms with $\hat H_0\,$ and 3 with $\hat W$, which compensate each other in a partly tricky way to zero.

Inserting the result~(\ref{eq:uccpt}) in Eq.~(\ref{eq:xak3}) for the 
$p$-$h$ amplitude $x_{ak}$ (and using $S^{(1)}_{ak} = t^{(1)}_{ak}$)  gives the simple expression 
\begin{equation}
x_{ak}^{(3)} = t_{ak}^{(3)} - \half I_0^{(2)} t_{ak}^{(1)}
\end{equation}
Noting that the PT expansion of $I_0$ is of the form $I_0 = 1 + I_0^{(2)} + \dots$ so that $I_0^{-1/2} = 1 - \half I_0^{(2)} + \dots$,
this is just the third-order level of the general relation 
\begin{equation}
x_{ak} = t_{ak} I_0^{-1/2}
\end{equation}
between the $p$-$h$ amplitudes in the UCC and BCC ground state accordingt to 
Eq.~(\ref{eq:ubcc}).

While the UCC and BCC ground states differ only by the normalization factor $I_0^{-1/2}$, the construction of the UCC ground state goes along with an enormous complication compared to the BCC procedure. This is seen for the first time in third order, where the 
UCC amplitudes $S_{ak}^{(3)}$ differ significantly from the BCC amplitudes $t_{ak}^{(3)}$ and their evaluation via the UCC amplitude
equations is already remarkably elaborate.
The complexity of the UCC formulation encountered in the simple NIP case may be seen as an indication that major 
challenges will arise in the UCC treatment of interacting particles, where the $\hat \sigma$ operator also comprises double and higher excitations.

\subsection{Excited states}

\begin{figure}
\begin{equation*}
\vcenterpdf{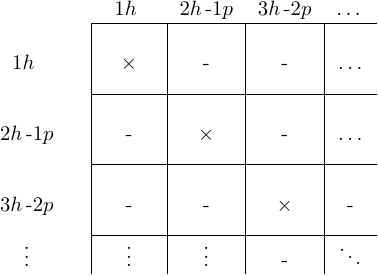}
\end{equation*}
\caption{Diagonal block structure of UCC and ISR secular matrices for non-interacting particles.}
\label{fig:corfigx}
\end{figure}

In analogy to the BCC approach discussed in Sec. II.B we will specifically focus on excitations in the ($N\!-\!1$)-particle system (particle removal). Here the UCC formulation is based on the 
basis set of states,
\begin{equation}
\label{eq:uccst}
\ket{\vtilde{\Psi}_K}=  e^{\hat \sigma}\hat{C}_{K} \ket{\Phi_0}
\end{equation}
where the $\hat \sigma$ operator is given by Eq.~(\ref{eq:siop})
and $\hat C_K$ denote the physical excitation operators of the manifold of $1h$, $2h$-$1p$, $\dots$, excitations as specified in Eq.~(\ref{eq:bccmf}).
Obviously, these states are orthonormal,
\begin{equation}
\braket{\vtilde{\Psi}_K}{\vtilde{\Psi}_L} = \delta_{KL}
\end{equation}
Accordingly, the representation of the (shifted) hamiltonian $\hat{H}-E^N_{0}$
with respect to these UCC states 
gives rise to a hermitian secular matrix, 
\begin{equation}
\label{eq:uccsm}
M^{ucc}_{IJ}=\bra{\vtilde{\Psi}_{I}}\hat{H}-E^N_{0}\ket{\vtilde{\Psi}_{J}}
= \dirint{\Phi_0}{\hat{C}_{I}^{\dagger} e^{-\hat{\sigma}}
\hat{H}  e^{\hat{\sigma}} \hat{C}_{J}}{\Phi_{0}} - E^N_{0}\delta_{IJ}
\end{equation}
By contrast to the BCC secular 
matrix~(\ref{eq:ccsm}), the UCC secular matrix $\bs M^{ucc}$ is block-diagonal (see Fig.~\ref{fig:corfigx}) with respect to the excitation classes $\nu = 1,2, \dots$: 
\begin{equation}
\label{eq:eq:bstruct2}
M^{ucc}_{IJ} = 0 \;\;\text{if}\;\; [I] \neq [J]
\end{equation}
Here and below, $[J]$ denotes the class of the configuration $J$, e.g., $[akl] = 2$ for a $2h$-$1p$ excitation. A proof of this property is given in Appendix B.
Of interest for the following are also the effective transition 
(or spectral) amplitudes (ETA)
\begin{equation}
\label{eq:uccsa}
\vtilde f_{Ip} = \dirint{\vtilde{\Psi}_{I}}{c_p}{\Psi_0} = \bra{\Phi_0} \hat C^\dagger_I e^{-\hat{\sigma}}
c_p  e^{\hat{\sigma}}\ket{\Phi_0}
\end{equation}
which are generally needed in the computation of spectral intensities.
Let us note that $\vtilde f_{Ip}$ vanishes unless $I$ refers to a $1h$ excitation ($[I] = 1$).

\subsection{Comparison with NIP-ADC/ISR}

The new approach to QDPT presented in \textbf{I} was based on the 
ADC/ISR states for the $1h$ excitations of $N\!-\!1$ non-interacting particles. These ADC/ISR states are identical with the NIP-UCC basis states~(\ref{eq:uccst}) just discussed. This remarkable finding shall be proven in the following.

Let us first briefly review the ISR procedure for the case of non-interacting particles. The starting point for the construction of the intermediate states is constituted by
the so-called correlated excited (CE) states. 
For the lowest class of states, that is, the $1h$ states, the CE 
precursors read
\begin{equation}
\label{eq:coexstat}
\ket{\Psi_{k}^{0}}=c_k \ket{\Psi_0}, \;\; k= 1, \dots, N
\end{equation}
from which the $1h$ intermediate states are obtained according to
\begin{equation}
\label{eq:isconstr2}
\ket{\tilde{\Psi}_k}=\sum_{l} \ket{\Psi^{0}_l} (\bs S^{-1/2})_{lk}
\end{equation}
as the result of symmetrical orthonormalization of the $1h$ CE states. Here 
\begin{equation}
\label{eq:cesolm}
S_{ij} = \braket{\Psi_{i}^{0}}{\Psi_{j}^{0}} = \dirint{\Psi_0}{c_i^\dagger c_j}{\Psi_0}, \;\; i,j \leq N 
\end{equation}
are the matrix elements of the CE state overlap matrix $\bs S$, which also can be seen as 
the $h/h$-block of the one-particle density matrix, $\bs S = \bs \rho_{hh}$  

The states of the next higher class, that is, the $2h$-$1p$ states, are constructed by a two-step procedure. First, precursor states are formed by Gram-Schmidt (GS) orthogonalization of the $2h$-$1p$ CE states, $\ket{\Psi^{0}_{akl}}=c_a^\dagger c_k c_l \ket{\Psi_0}$, with regard to the already constructed $1h$ intermediate states:  
\begin{equation}
\label{eq:isconstr1}
\ket{\Psi^{\#}_{akl}}=c_a^\dagger c_k c_l \ket{\Psi_0}- \sum_{j}\ket{\tilde{\Psi}_j}
\dirint{\tilde{\Psi}_j}{c_a^\dagger c_k c_l}{\Psi_0}
\end{equation}
In the second step, again symmetrical orthonormalization is applied to the $2h$-$1p$ precursor states, yielding the  
$2h$-$1p$ intermediate states, $\ket{\tilde{\Psi}_{akl}}$.
In an analogous way, the intermediate states of higher excitation classes,
$3h$-$2p$, $4h$-$3p, \dots$, can successively be constructed. 

By construction, the ISR states are orthonormal,
\begin{equation}
\braket{\tilde{\Psi}_K}{\tilde{\Psi}_L} = \delta_{KL}
\end{equation}
and the ADC/ISR secular matrix, $\bs M$, obtained according to
\begin{equation}
\label{eq:isrsm}
M_{IJ}=\bra{\tilde{\Psi}_{I}}\hat{H}-E^N_{0}\ket{\tilde{\Psi}_{J}}
\end{equation}
is hermitian.
The corresponding ETA elements are given by
\begin{equation}
\label{eq:isrsa}
 f_{Ip} = \dirint{\tilde{\Psi}_{I}}{c_p}{\Psi_0} 
\end{equation}
For occupied orbitals, $p \leq N$, $f_{Ip}$ vanishes if $[I] >  1$.

Like the UCC secular matrix, $\bs M$ is block-diagonal (see Fig.~\ref{fig:corfigx}) with respect to the excitation classes,   
\begin{equation}
\label{eq:eq:bstruct3}
M_{IJ} = 0 \;\;\text{if}\;\; [I] \neq [J]
\end{equation}
A proof of this property, taking recourse to the block structure of the NIP-BCC secular matrix $\bs M^{bcc}$, was given in \textbf{I}. 
Alternatively, the block-diagonality of $\bs M$ can also be shown with the help of the NIP-UCC representation, as 
is briefly addressed at the end of this section.

To show that the ISR and UCC $1h$-states are identical, 
$\ket{\tilde{\Psi}_k} = \ket{\vtilde{\Psi}_k}$, we first consider 
the $1h$-$h$ blocks of the respective ETA matrices, $\bs f_{11}$ and  
$\vtilde{\bs f}_{11}$. 
According to Eq.~(\ref{eq:isrsa}), the matrix elements of 
$\bs f_{11}$ read
\begin{equation}
\label{eq:isrsa11}
 f_{kl} = \dirint{\tilde{\Psi}_{k}}{c_l}{\Psi_0}, \;k,l \leq N 
\end{equation}
Following the explicit construction of $\ket{\tilde{\Psi}_{k}}$ according to Eq.~(\ref{eq:isconstr2}) the further evaluation yields
\begin{equation}
\label{eq:saexpl}
 f_{kl} = \sum_i (S^{-1/2})^*_{ik} \dirint{\Psi_0}{c_i^\dagger c_l}{\Psi_0}
 = \sum_i (S^{-1/2})_{ki} S_{il} = (S^{-1/2})_{kl}
\end{equation}
where the hermiticity of the CES overlap matrix $\bs S$ (Eq.~\ref{eq:cesolm}) has been used.
In matrix form this result can compactly be written as
\begin{equation}
\label{eq:f11}
\bs f_{11} = \bs S^{1/2}
\end{equation}
As discussed in \textbf{I}, $\bs S$ can be expressed in terms of the 
$11$-block of the one-particle eigenvector matrix $\bs X$ (see Eq.~\ref{eq:part0}):
\begin{equation}
\bs{S} =  (\bs X_{11} \bs  X_{11}^\dagger)^t
\end{equation}

The UCC counterpart to $\bs f_{11}$ is the $\vtilde{\bs{f}}_{11}$ block of the amplitudes 
\begin{equation}
\label{eq:fkl}
\vtilde{f}_{kl} = \dirint{\vtilde{\Psi}_k}{c_l}{\Psi_0} =
\dirint{\Phi_0}{c^\dagger_k\, e^{-\hat \sigma} c_l\, e^{\hat \sigma}}{\Phi_0}, \; k,l \leq N
\end{equation}
In fact, $\bs f_{11}$ and $\vtilde{\bs{f}}_{11}$ are identical, as we will show in the following.

Let us consider the overlap matrix~(\ref{eq:cesolm})
of the CE states $\ket{\Psi_j^0}$,
\begin{equation} 
 S_{kl} = \dirint{\Psi_0}{c_k^\dagger c_l}{\Psi_0}, \; k,l \leq N
\end{equation}
and insert on the right-hand side the resolution-of-identity (ROI) in terms of the UCC states, 
\begin{equation}
\hat{\mathbb{1}} = \sum_I \ket{\vtilde{\Psi}_I} \bra{\vtilde{\Psi}_I}
\end{equation}
Since 
$\dirint{\Phi_0}{C^\dagger_I e^{-\hat \sigma} c_l\, e^{\hat \sigma}}{\Phi_0}$
vanishes if the rank $[I]-1$ of $C^\dagger_I$  is larger than zero (see App.~B), that is, for 
the excitation classes $2,3,\dots $ 
the ROI insertion can be truncated after the $1h$ excitations. 
Accordingly, we obtain 
\begin{equation}
 S_{kl} = \sum_i \dirint{\Psi_0}{c_k^\dagger}{\vtilde{\Psi}_i}
\dirint{\vtilde{\Psi}_i}{c_l}{\Psi_0} = \sum_i \vtilde{f}_{ik}^*
\vtilde{f}_{il}
\end{equation}
where the definition~(\ref{eq:fkl}) of the UCC amplitudes has been used in the second equation.
In matrix form, this result can be stated as
\begin{equation}
\bs{S} = \vtilde{\bs{f}}_{11}^\dagger \vtilde{\bs{f}}_{11}
\end{equation}
Now, using that $\vtilde{\bs{f}}_{11}$ is hermitian, which will be proven separately below, 
yields $\bs{\mathsf{S}} = \vtilde{\bs{f}}_{11}^2$, 
and, finally,  
\begin{equation}
\label{eq:fs}
\vtilde{\bs{f}}_{11} = \bs{S}^{1/2} = \bs{f}_{11}
\end{equation}
(Note that the definiteness of the square root of a matrix supposes that the matrix is positive definite.)\\
\\
\textbf{Proof of the hermiticity of $\vtilde{\bs{f}}_{11}$}:\\
Here we start from using the BCH expansion
\begin{equation}
\label{eq:bch}
e^{-\hat \sigma} c_l\, e^{\hat \sigma} =  c_l +[c_l, \hat \sigma] + \half [[c_l, \hat \sigma],\hat \sigma]] + \dots
\end{equation}
in Eq.~(\ref{eq:fkl}), where $\hat \sigma = \hat S - \hat S^\dagger$ is composed of a physical and an unphysical operator.
The nested commutators arising here can successively be evaluated as follows.
Being a physical operator, $c_l$ commutes with $\hat S$. The commutator with 
$\hat S^\dagger$ results in a sum of unphysical operators (of rank 0), which in turn commute with  $\hat S^\dagger$  but not with $\hat S$. This can be continued further. 
Let us write the BCH expansion~(\ref{eq:bch}) in the form
 \begin{equation}
e^{-\hat \sigma} c_l\, e^{\hat \sigma} =\hat K_l^0 + \hat K_l^1 + \half \hat K_l^2 + \frac{1}{6} \hat K_l^3 + \dots
\end{equation}
where $ \hat K_l^n$ denotes the $n$-fold commutator.

The nested commutators can successively be evaluated as shown below up to $n=4$: 
\begin{align}
\nonumber
\hat K_l^0 &= c_l \\
\nonumber
\hat K_l^1 & = -[c_l, \hat S^\dagger] = -[\hat K_l^0 ,\hat S^\dagger] = - \sum_b S_{bl}^*\,c_b \\
\nonumber
\hat K_l^2 & = -[[c_l, \hat S^\dagger],\hat S] = [\hat K_l^1,\hat S] =
- \sum_{b,i} S_{bl}^*S_{bi}\,c_i \\
\nonumber
\hat K_l^3 & = [[[c_l, \hat S^\dagger],\hat S],\hat S^\dagger] = -[\hat K_l^2,\hat S^\dagger] =
\sum_{b,i,c} S_{bl}^*S_{bi}S_{ci}^*\,c_c \\
\nonumber
\hat K_l^4 & = [[[[c_l, \hat S^\dagger],\hat S],\hat S^\dagger],\hat S] = [\hat K_l^3,\hat S] =
 \sum_{b,i,c,j} S_{bl}^*S_{bi}S_{ci}^*S_{cj}\,c_j \\
\label{eq:nestcom}
&\vdots
\end{align}
It is obvious how this procedure can be continued or even formulated as a general iterative scheme.
Each nested commutator $\hat K^n_l$ results in an explicit sum of fermion operators of rank zero, $c_p$, 
\begin{equation}
\hat K^n_l = \sum_p z^{(n)}_{lp} c_p
\end{equation}
being physical operators, $p \leq N$, for even $n$ and unphysical operators, $p > N$, for odd $n$.  
The coefficients $z^{(n)}_{lp}$ can be obtained successively as products of the 
UCC amplitudes $S_{ak}$ and $S^*_{ak}$. 
Note that the overall sign corresponds to the number of the respective $-\hat S^\dagger$ operators.

What are the contributions of the nested commutators to Eq.~(\ref{eq:fkl})? Here we consider the matrix elements
\begin{equation}
\dirint{\Phi_0}{c_k^\dagger \hat K^n_l}{\Phi_0} = 
\begin{cases}
&z^{(n)}_{lk}, \; n \;\text{even}\\
&0, \; n \; \text{odd}
\end{cases}
\end{equation} 
That is, only $\hat K^n_l$ with even $n$ come into play, and their respective 
contribution is the coefficient $z^{(n)}_{lk}$ of the operator $c_k$.  
In the case $n=4$, for example, this coefficient is given by
\begin{equation}
z^{(4)}_{lk} = \sum_{b,i,c} S_{bl}^*S_{bi}S_{ci}^*S_{ck}
\end{equation}
As is readily seen, these coefficients are hermitian,
\begin{equation}
z^{(4)*}_{kl} = z^{(4)}_{lk}
\end{equation}
and, as the inspection of Eqs.~(\ref{eq:nestcom}) shows, this property applies to any $n$-fold nested commutator with even $n$. 
Thus, we may conclude that $\vtilde{\bs f}_{11}$ is a hermitian matrix.\\

The identity of the ISR and UCC effective transition amplitudes, $\bs{f}_{11}=\vtilde{\bs{f}}_{11} = \bs{S}^{1/2}$, suggests an even more stringent proposition, namely the identity of the ISR and UCC $1h$ states themselves,
$\ket{\vtilde{\Psi}_k} = \ket{\tilde{\Psi}_k}$.
To show this we consider the matrix elements
\begin{equation}
U_{kl} = \braket{\vtilde{\Psi}_k}{\tilde{\Psi}_l}, \; k,l \leq N
\end{equation}
of the unitary transformation $\bs U$, relating the UCC and ISR states.
Using Eq.~(\ref{eq:isconstr2}) for the construction of the ISR states, 
Eqs.~(\ref{eq:fkl}) and (\ref{eq:fs}) for the UCC spectral amplitudes  
the $U_{kl}$ matrix elements can readily evaluated 
\begin{equation}
\braket{\vtilde{\Psi}_k}{\tilde{\Psi}_l}
 = \sum_i \braket{\vtilde{\Psi}_k}{\Psi^0_i}(S^{-1/2})_{il}
= \sum_i \vtilde f_{ki}(S^{-1/2})_{il} 
= \sum_i (S^{1/2})_{ki}(S^{-1/2})_{il} = \delta_{kl}
\end{equation} 
yielding the expected result. In matrix form this can be written as
\begin{equation}
\bs U_{11} = \bs 1
\end{equation}
where $\bs U_{11}$ denotes the $1h$-$1h$ block of $\bs U$.

One may wonder whether the identity of the $1h$ ISR and UCC states also applies to the states of higher excitation classes. This question is addressed in App. C. There, specifically, it is established that the unitary transformation matrix $\bs U$ is block-diagonal with respect to the excitation classes, which means that the ISR and UCC states of higher classes are essentially equivalent.

The unitary transformation $\bs U$ establishes the relation
\begin{equation}
\bs M = \bs U^\dagger \bs M^{ucc} \bs U
\end{equation}
between the ISR and UCC secular matrices.
Since both $\bs M^{ucc}$ and $\bs U$ are block-diagonal, as shown in App.~B and App.~C, respectively, the latter equation implies that the ISR secular matrix, $\bs M$, is block-diagonal as well.

\section{Concluding remarks}

In extension to a recent paper~\cite{sch23:1}, we have studied in this article the 
application of the familiar biorthogonal (BCC) and unitary (UCC) coupled-cluster methods to systems of non-interacting particles (NIP). To emphasize it once again, the physics of NIP-systems is trivial. Nevertheless, the NIP simplifications of actual many-body methods can be useful to analyze and clarify methodological aspects of the respective approach. 

With regard to the ground state, the NIP versions illustrate the enormous complexity of the UCC versus the BCC treatment. While in both cases the cluster operator in the exponential comprises only single ($p$-$h$) excitations, the BCC amplitude equations terminate after the quadratic term in the expansion of the exponential CC operator, and there is even a closed-form expression for the BCC amplitudes in terms of eigenvector components of the underlying one-particle system. The UCC amplitude equations, by contrast, do not terminate after a finite number of terms featuring intricate mutual compensations of the physical (excitation) and unphysical (de-excitation) parts in the cluster operator. To a certain extent, the complexity of the UCC amplitude equations becomes apparent by using perturbation theory in their solution, as was demonstrated in Sec.~III.A by expanding the UCC amplitudes through third order.  

The BCC and UCC treatment of excitations, here specifically of 
$N\!-\!1$ particles, is based on a representation of the (shifted) hamiltonian in terms of specific CC states, which build on the respective CC ground-state. 
In the NIP case, the resulting hermitian UCC secular matrix is block-diagonal with respect to the classes of $1h$, $2h$-$1p$, $\dots$, excitations. The non-hermitian BCC secular matrix, on the other hand, exhibits a hybrid structure, being block-diagonal in the lower left part and featuring coupling blocks between successive excitation classes in the upper right part. The block structure in the upper right part reflects the fact that the bi-orthogonal CC states used as the left expansion manifold are essentially of CI-type~\cite{sch09:145}. 

There is a relationship between the ISR, BCC, and UCC approaches in that the states in their respective expansion manifolds derive from the correlated ground state rather than the "unperturbed" (Hartree-Fock) ground state used in the CI treatment. In the BCC scheme, the right-hand expansion manifold is formed by CE states, $\hat C_I\ket{\Psi_0}$, with the BCC parametrization of the ground state. The ISR expansion manifold is obtained from the CE states via a specific orthonormalization procedure. And the UCC expansion states can be seen as resulting by applying UCC transformed excitation operators to the correlated ground state. The interesting finding is that in the NIP case the ISR and UCC states are essentially equivalent, being reflected by the block-diagonality of the unitary transformation relating the UCC to the ISR states. They may 
even be identical, which so far, however, could be established only for the lowest excitation class.

For specificity, we here have focussed on the excitations of $N\!-\!1$ non-interacting particles (detachment). However, it should be clear that the treatment can readily be transferred to other cases as well, such as $N$-particle (neutral) excitations, ($N\!+\!1$)-particle (attachment) excitations , or even attachment and detachment of two (or more) particles.

\newpage
\appendix
\renewcommand{\theequation}{A.\arabic{equation}}
\setcounter{equation}{0}
\section*{Appendix A: Thouless solution and the BCC amplitude equations}

In the following we want to show that the Thouless expression~(\ref{eq:t21}) for the $t_{ak}$ amplitudes, 
\begin{equation}
\bs t_{21} = \bs X_{21} \bs X_{11}^{-1}
\end{equation}
is indeed a solution of the BCC amplitude equations Eqs.~(\ref{eq:bccgs})

Let us first write Eqs.~(\ref{eq:bccgs}) in matrix form,
\begin{equation}
\label{eq:bccmat}
- \tilde{\bs t}_{21} = \bs W_{21} + \bs W_{22}\, \bs t_{21} - \bs t_{21}\bs W_{11} - \bs t_{21}\bs W_ {12}\,\bs t_{21}
 \end{equation}
where $\tilde{\bs t}_{21}$ denotes the matrix of modified amplitudes,
\begin{equation}
\tilde t_{ak} = t_{ak} (\epsilon_a - \epsilon_k)
\end{equation}
Next, we insert the Thouless expression on the right-hand side and 
multiply by $\bs 1 = \bs X_{11}\bs X_{11}^{-1}$:
\begin{align}
\nonumber
- \tilde{\bs t}_{21} \stackrel{!}{=} & \;\bs W_{21} + \bs W_{22} \bs X_{21} \bs X_{11}^{-1} - \bs X_{21} \bs X_{11}^{-1}\bs W_{11} 
- \bs X_{21} \bs X_{11}^{-1}\bs W_{12}\bs X_{21} \bs X_{11}^{-1}\\
\label{eq:zwr}
=& \left \{(\bs W_{21}\bs X_{11} + \bs W_{22} \bs X_{21}) - \bs X_{21} \bs X_{11}^{-1}(\bs W_{11} \bs X_{11} +
\bs W_{12}\bs X_{21} )\right \} \bs X_{11}^{-1}
\end{align}
where of course the validity of the equality sign is still to be shown. 
To proceed we consider the secular equation~(\ref{eq:ase}) in the 
partitioned matrix form:
\begin{equation}
\left(
\begin{array}{cc}
\bs \epsilon_1 + \bs W_{11} & \bs W_{12} \\
\bs W_{21} &\bs \epsilon_2 + \bs W_{22}
\end{array}
\right) \bs X = 
\bs X 
\left( \begin{array}{cc}
\bs E_1 & \bs 0 \\
\bs 0 &  \bs E_2
\end{array}
\right )
\end{equation} 
Using the parttioning of the eigenvector matrix $\bs X$ according to 
Eq.~(\ref{eq:part0}), one obtains the following relations for the upper left block ($\bs{11}$) and the lower left block ($\bs{21}$): 
\begin{align}
\bs \epsilon_1 \bs X_{11} + (\bs W_{11}\bs X_{11} + \bs W_{12} \bs X_{21}) &= \bs X_{11}\bs E_1\\
\bs \epsilon_2 \bs X_{21} + (\bs W_{21}\bs X_{11} + \bs W_{22} \bs X_{21}) &= \bs X_{21}\bs E_1
\end{align}
The two terms in round brackets can be retrieved in the right-hand side of Eq.~(\ref{eq:zwr}). This allows for corresponding replacements yielding  
\begin{align*}
- \tilde{\bs t}_{21} =& \left \{\bs X_{21}\bs E_1 - \bs\epsilon_2 \bs X_{21} - \bs X_{21} \bs X_{11}^{-1}
(\bs X_{11} \bs E_1 - \bs \epsilon_1 \bs X_{11}) \right \} \bs X_{11}^{-1}\\
=&  -\bs\epsilon_2 \bs X_{21}\bs X_{11}^{-1}  
+ \bs X_{21}\bs X_{11}^{-1} \bs\epsilon_1 
\end{align*}
The final result is
\begin{equation}
\tilde{\bs t}_{21} =   \bs\epsilon_2 \bs t_{21}
- \bs t_{21}\bs\epsilon_1 
\end{equation}
This is indeed an identity as the multiplication of $\bs t_{21}$ from the left by $\bs \epsilon_2$ minus the result of multiplicating from the right by 
$\bs \epsilon_1$ is exactly the operation that transfers $\bs t_{21}$ into 
$\tilde{\bs t}_{21}$.

In a similar way it can be shown that the BCC expression for the ground-state energy,
\begin{equation}
E_0^{CC} = \dirint{\Phi_0}{(\hat H_0 + \hat W) e^{\hat T_1}}{\Phi_0} = \sum_k (\epsilon_k + w_{kk}) + \sum_{b,l}
t_{bl}\,w_{lb} 
\end{equation}
with the amplitudes given by $\bs t_{21} = \bs X_{21} \bs X_{11}^{-1}$ 
reproduces the original expression
\begin{equation}
E_0 = \dirint{\Psi_0}{\hat H}{\Psi_0} = \sum_{n=1}^N e_n = tr(\bs E_1)  
\end{equation}
for the ground-state energy of the NIP system.

\newpage
\renewcommand{\theequation}{B.\arabic{equation}}
\setcounter{equation}{0}
\section*{Appendix B: Block-diagonal structure of the UCC secular matrix}

The non-diagonal elements of the NIP-UCC secular matrix~(\ref{eq:uccsm}) are given by 
\begin{equation}
M^{ucc}_{IJ}= \dirint{\Phi_0}{\hat{C}_{I}^{\dagger} e^{-\hat{\sigma}}
\hat{H}  e^{\hat{\sigma}} \hat{C}_{J}}{\Phi_{0}}, \; I\neq J
\end{equation}
where $\hat \sigma$ is the NIP-UCC excitation operator~(\ref{eq:siop}).
Via the Baker-Campbell-Hausdorff (BCH) expansion the unitary-transformed hamiltonian (or its constituents 
$\hat H_0$ and $\hat W$) can be expressed according to 
\begin{equation*}
e^{-\hat{\sigma}} \hat H e^{\hat{\sigma}} = \hat H + [\hat H, \hat \sigma] + 
\half [[\hat H, \hat \sigma], \hat \sigma] + \dots
\end{equation*}
Both $\hat H$ and $\hat \sigma$ are operators of rank 1, and the commutator of two operators of rank 1 is again an operator of rank 1. (The rank of a fermion operator product denotes the number of creation operators; the rank of a general operator is given by the maximal rank of its operator product constituents.)
Accordingly, each of the nested commutators in the BCH expansion is of rank 1, which shows that the transformed hamiltonian $e^{-\hat{\sigma}} \hat{H} e^{\hat{\sigma}}$ is itself an operator of rank 1.

Now, a matrix element of the type  
\begin{equation*}
\dirint{\Phi_{0}}{\hat{C}_{I}^{\dagger}\hat O(1) \hat{C}_{J}}{\Phi_{0}} 
\end{equation*}
whith $\hat O(1)$ being an operator of rank 1, vanishes if $[I] > [J] + 1$ or $[J] > [I] + 1$. 
Note that 
the rank of an ($N\!-\!1$)-particle excitation operator $\hat C_K$ is given by $[K]-1$.
  
So it remains to show the assertion for $[I] = [J] \pm 1$.
Let us consider the case 
where $I = akl$ is a $2h$-$1p$ excitation (class 2) and  $J = j$ is a $1h$ excitation (class 1), 
\begin{equation}
\label{eq:maklj}
M^{ucc}_{akl,j} = \dirint{\Phi_{0}}{c_l^\dagger c_k^\dagger c_a e^{-\hat{\sigma}}
\hat H e^{\hat{\sigma}} c_j}{\Phi_{0}}
\end{equation}

As we have seen, the unitary transformed hamiltonian is a (particle number conserving) operator of rank 1, 
so that it can be written in the general form
\begin{equation}
\label{eq:zuv}
 e^{-\hat{\sigma}}\hat H e^{\hat{\sigma}} = \sum_{u,v} Z_{uv} c_u^\dagger c_v
\end{equation}
where $Z_{uv}$ are coefficients depending on the orbital energies, 
$\epsilon_p$, the matrix elements, $w_{pq}$, and increasing powers of the UCC amplitudes, $S_{bl}$, and their complex conjugates. 
Using this form in Eq.~(\ref{eq:maklj})  yields the simple expression
\begin{equation}
M^{ucc}_{akl,j} = \delta_{lj} Z_{ak} - \delta_{kj} Z_{al}
\end{equation}
which, however, vanishes because the coefficients $Z_{ak}, Z_{al}$ vanish as a result of the UCC amplitude equations~(\ref{eq:uccamp}). 
This can be seen by using Eq.~(\ref{eq:zuv}) in the latter equations,   
\begin{equation}
0 = \dirint{\Phi_{ak}}{e^{-\hat \sigma} \hat H e^{\hat \sigma}}{\Phi_0} = Z_{ak},  \; a > N, k \leq N
\end{equation}
The amplitude equations ensure that the coefficients $Z_{ak}$
of the physical operators $c_a^\dagger c_k, a > N, k \leq N$ in the 
expansion~(\ref{eq:zuv}) of the transformed hamiltonian vanish, or, stated differently, there are no physical excitation operators in $e^{-\hat{\sigma}}\hat H e^{\hat{\sigma}}$. 

\newpage
\renewcommand{\theequation}{C.\arabic{equation}}
\setcounter{equation}{0}
\section*{Appendix C: Block-diagonality of the unitary transformation between ISR and UCC states of a NIP system}

In Sec.~III.C it was shown that the $1h$ ISR and UCC states are identical. In this Appendix we address the question whether such a finding applies as well to the states of higher excitation classes.
Here, we establish that the unitary transformation relating the   
UCC and ISR states is block-diagonal with regard to the excitation classes.  

The unitary transformation $\bs U$ relates the UCC and ISR states  
of $N$-$1$ particles according to
\begin{equation}
\label{eq:uij}
U_{IJ} = \braket{\vtilde{\Psi}_I}{\tilde{\Psi}_J}
\end{equation}
First we show that $\bs U$ is block-diagonal with respect to the excitation classes in the lower left part:
\begin{equation}
\label{eq:pbdu}
U_{IJ} = 0 \;\; \text{for} \;\; [I] > [J]
\end{equation}
Here, as before, $[K]$ denotes the excitation class of the configuration $K$.

According to the ISR construction, as recapitulated in Sec.~III.C, the state 
$\ket{\tilde{\Psi}_J}$ of class $[J]$ is given as a linear combinations of CE states of the class $[J]$ and lower classes:
\begin{equation}
\ket{\tilde{\Psi}_J} = \sum_{K,[K]\leq[J]} z_K \hat C_K \ket{\Psi_0}
\end{equation}
Using this expression and the explicit form
\begin{equation}
\ket{\vtilde{\Psi}_I} = e^{\hat \sigma} \hat C_I \ket{\Phi_0}
\end{equation}
of the UCC state $\ket{\vtilde{\Psi}_I}$, Eq.~(\ref{eq:uij})
can be written as
\begin{equation}
\label{eq:uijx}
U_{IJ} = \sum_{K,[K]\leq[J]} z_K \dirint{\Phi_0}{\hat C_I^\dagger
e^{-\hat \sigma}\hat C_K e^{\hat \sigma}}{\Phi_0}
\end{equation}
As in App.~B, the BCH expansion can be used to analyze the operators $e^{-\hat \sigma}\hat C_K e^{\hat \sigma}$ on the right-hand side of the latter equation. Since $\hat \sigma$ is an operator of rank $1$, the UCC transformation does not increase the rank of the original excitation operator $\hat C_K$, being $[K] - 1$. 
This means that the matrix elements on the right-hand side of Eq.~(\ref{eq:uijx}) are of the type $\dirint{\Phi_0}{\hat C_I^\dagger \hat O_K}{\Phi_0}$, where $\hat O_K$  is an operator of rank $[K]-1$. Since the rank $[K]-1$ is smaller than the rank of $\hat C_J$ and, thus, by assumption smaller than the rank of $\hat C_I$, that is, $[K]-1  < [I] -1$, all matrix elements necessarily vanish, which proves  Eq.~(\ref{eq:pbdu}).  

As a consequence of the unitarity of $\bs U$, $\bs U^\dagger \bs U = \bs 1$, the block-diagonality of the lower left part implies block-diagonality for the entire matrix:
\begin{equation}
\label{eq:bdofu}
U_{IJ} = 0 \;\;\text{for} \;\; [I] \neq [J]
\end{equation}
This can be shown as follows.
The columns of $\bs U$ are orthonormal. Let $\dul U_{\,\nu}$ denote
the block of columns associated with the excitation class $\nu$, 
\begin{equation}
\dul U_{\,\nu} = \left(
\begin{array}{c}
\bs U_{1\nu} \\
\bs U_{2\nu} \\
\vdots
\end{array}
\right)
\end{equation}
The orthonormaliy of $\dul U_{\,1}$,
\begin{equation}
\dul U_{\,1} = \left(
\begin{array}{c}
\bs U_{11} \\
\bs 0 \\
\vdots
\end{array}
\right)
\end{equation}
implies that the 11-block $\bs U_{11}$ is unitary. Now, the columns of 
$\dul U_{\,1}$ are orthogonal to those of $\dul U_{\,2}$, which 
translates into the matrix equation
\begin{equation}
\dul U_{\,1}^\dagger \, \dul U_{\,2} = \bs U_{11}^\dagger \,\bs U_{12} = \bs 0
\end{equation}
Using that $\bs U_{11}$ is unitary, we may conclude  
$\bs U_{12} = \bs 0$.
In the same way, the orthogonality of $\dul U_{\,1}$ and $\dul U_{\,\nu}$ for $\nu  > 2$ requires that the respective first blocks vanish:
\begin{equation}
\bs U_{1 \nu} = \bs 0, \;\; \nu \geq 2
\end{equation}
Having established that $\dul U_{\,2}$ is of the form
\begin{equation}
\dul U_{\,2} = \left(
\begin{array}{c}
\bs 0 \\
\bs U_{22} \\
\bs 0 \\
\vdots
\end{array}
\right)
\end{equation}
where $\bs U_{22}$ is unitary, 
the orthogonality of $\dul U_{\,2}$ and $\dul U_{\,\nu}$ for $\nu =3,4,\dots$ leads to the equations
\begin{equation}
\bs U_{22}^\dagger \,\bs U_{2 \nu} = \bs 0, \; \; \nu = 3,4, \dots 
\end{equation} 
In view of the unitarity of $\bs U_{22}$, this in turn ensures
that 
\begin{equation}
\bs U_{2\,\nu} = \bs 0, \; \;\nu \geq 3
\end{equation}

Obviously, this procedure can be carried on and readily be cast into a formally correct proof of the block-diagonal structure in the upper right part of $\bs U$. We note that a similar proof scheme
has been used in establishing the so-called canonical order relations in unitary transformation matrices (see App. C in Ref.~\cite{mer96:2140}).

The block-diagonality of $\bs U$ means that the UCC and ISR states of a given excitation class $\nu$ are essentially equivalent, since they differ at most by a unitary transformation within that class, as given by the corresponding block $\bs U_{\nu\nu}$. But is it possible to go beyond that and even establish their identiy as for the $1h$ states? This will be briefly discussed in the following, considering here specifically the case of the $2h$-$1p$ excitations.

The $2h$-$1p$ IRS states $\tilde{\Psi}_{akl}, k<l,$ derive from the precursor states,
\begin{equation}
\label{eq:aklpcs} 
\ket{\Psi^{\#}_{akl}}= \ket{\Psi^0_{akl}}- \sum_{j}\ket{\tilde{\Psi}_j}
\braket{\tilde{\Psi}_j}{\Psi^0_{akl}}
\end{equation}
via symmetrical orthonormalization based on the overlap matrix
$\lo{\bs S}$ with the elements
\begin{equation}
\lo S_{akl,a'k'l'} = \braket{\Psi^\#_{akl}}{\Psi^\#_{a'k'l'}}
=\braket{\Psi^0_{akl}}{\Psi^0_{a'k'l'}} - \sum_j \braket{\Psi^0_{akl}}{\tilde{\Psi}_j} \braket{\tilde{\Psi}_j}{\Psi^0_{a'k'l'}}
\end{equation}
As in Sec.~III.C, we consider the ETA amplitudes associated with the
$2h$-$1p$ ISR states
\begin{equation}
\label{eq:f22}
f_{akl,a'k'l'} = \braket{\tilde{\Psi}_{akl}}{\Psi^0_{a'k'l'}} =
\braket{\tilde{\Psi}_{akl}}{\Psi^\#_{a'k'l'}}
\end{equation}
Here, replacing the CE state $\ket{\Psi^0_{a'k'l'}}$ with the precursor state $\ket{\Psi^\#_{a'k'l'}}$ in the second equation is justified since
$\tilde{\Psi}_{akl}$ is orthogonal to the $1h$ states $\tilde{\Psi}_{j}$.
In analogy to the derivation of Eq.~(\ref{eq:f11}), the relation 
\begin{equation}
\label{eq:f22S}
\bs f_{22} = \lo{\bs S}^{1/2}
\end{equation}
can readily be established, where $\bs f_{22}$ denotes the matrix of the  
ETA elements in Eq.~(\ref{eq:f22}).

For the $2h$-$1p$ UCC states the corresponding ETA matrix 
$\vtilde{\bs f}_{22}$ is constituted by the elements
\begin{equation}
\vtilde f_{akl,a'k'l'} = \braket{\vtilde{\Psi}_{akl}}{\Psi^0_{a'k'l'}} =
\braket{\vtilde{\Psi}_{akl}}{\Psi^\#_{a'k'l'}}
\end{equation}
where the orthogonality of $\vtilde{\Psi}_{akl}$ and $\tilde{\Psi}_{j}$ has been used in the second equation.

Proceeding like in Sec.~III.C, that is, inserting the ROI in terms of the UCC states in the matrix elements of $\lo{\bs S}$, yields
\begin{equation}
\lo S_{akl,a'k'l'} = \sum_{b,i<j} \braket{\Psi^\#_{akl}}{\vtilde{\Psi}_{bij}} \braket{\vtilde{\Psi}_{bij}}{\Psi^\#_{a'k'l'}}
= \sum_{b,i<j} \vtilde f^*_{bij,akl} \vtilde f_{bij,a'k'l'} 
\end{equation}
where again obvious orthogonality conditions ensure that the ROI expansion can be restricted to the $2h$-$1p$ UCC states.
In matrix form, this can be written as 
\begin{equation}
\label{eq:Sff}
\lo{\bs S} = \vtilde{\bs f}^\dagger_{22} \vtilde{\bs f}_{22} 
\end{equation}
Finally, we consider the $\bs U_{22}$ block of the unitary transformation $\bs U$, featuring the matrix elements
\begin{equation}
U_{akl,a'k'l'} = \braket{\vtilde{\Psi}_{akl}}{\tilde{\Psi}_{a'k'l'}}
\end{equation}
Using the explicit construction of the $2h$-$1p$ ISR states, one obtains the following expression:
\begin{equation}
\bs U_{22} = \vtilde{\bs f}_{22} \lo{\bs S}^{-1/2}
\end{equation}
Obviously, $\bs U_{22}$ itself is unitary, but not necessarily a unit operator, $\bs U_{22} = \bs 1$. The latter would follow from
$\vtilde{\bs f}_{22} = \lo{\bs S}^{1/2}$, being itself a consequence of Eq.~(\ref{eq:Sff}) if $\vtilde{\bs f}_{22}$ were hermitian. 
So the missing link is the hermiticity of $\vtilde{\bs f}_{22}$. 
Unfortunately, we must leave this issue open. In fact, a procedure as the one used in Sec.~III.C for establishing the hermiticity of 
$\vtilde{\bs f}_{11}$ cannot simply be transferred to the matrix elements of $\vtilde{\bs f}_{22}$,
\begin{equation}
\vtilde f_{akl,a'k'l'} = \braket{\vtilde{\Psi}_{akl}}{\Psi^0_{a'k'l'}} 
= \dirint{\Phi_0}{c_l^\dagger c_k^\dagger c_a e^{-\sigma} c_{a'}^\dagger c_{k'} c_{l'}e^{-\sigma}}{\Phi_0}
\end{equation} 
as here the corresponding BCH expansion involves an operator product, $c_{a'}^\dagger c_{k'} c_{l'}$, for which the handling of nested commutators becomes cumbersome.

\end{document}